\documentclass{ws-procs9x6}
\newcommand{\sub}[1]{\ensuremath{_{\mathrm{#1}}}}
\newcommand{\pom}{\ensuremath{\mathbb{P}}}
\newcommand{\OurAlphaZero}{1.093 \pm 0.003 (stat.) ^{+0.008}_{-0.007} (syst.)}
\newcommand{\OurAlphaPrime}{0.116 \pm 0.027 (stat.) ^{+0.036}_{-0.046} (syst.)}
\newcommand{\GeV}{\ensuremath{\mathrm{GeV}}}
\newcommand{\der}{\ensuremath{\mathrm{d}}}

\newcommand{\super}[1]{\ensuremath{^{\mathrm{#1}}}}
\newcommand{\rhonull}{\ensuremath{\rho^{\circ}}\,}

\begin{document}

\title{A New Measurement of Exclusive \boldmath$\rho^{\circ}$~Photoproduction 
at HERA}

\author{J. Olsson\footnote{\uppercase{T}alk given at \uppercase{DIS} 2006, 
\uppercase{T}sukuba, \uppercase{J}apan,
 on behalf of the \   
{\bf \uppercase{H}1 \uppercase{C}ollaboration}.}}

\address{DESY, Hamburg, Germany \\
E-mail: jan.olsson@desy.de}
 
\maketitle

\abstracts{Exclusive \rhonull photoproduction has been measured with
high statistics by the H1 Collaboration. The kinematical range is 
$20 < W < 90$ GeV, $|t|<3$~GeV$^2$ and $Q^2<4$~GeV$^2$. 
Cross sections are measured 
differentially in $W$ and $t$ and allow the extraction of 
the Pomeron trajectory, for the first time
using \rhonull photoproduction data of one
single experiment. The preliminary result,
 $\alpha\sub\pom\,(t) = \OurAlphaZero + (\OurAlphaPrime)\,\GeV^{-2} \cdot t$,
deviates significantly from the canonical expectation.
}

\section{Introduction}

  Exclusive photoproduction of \rhonull 
  has been extensively studied in the last 40 years, first in
   fixed target experiments\cite{bib:rho-fixedtarget}, 
   and more recently in the $ep$ collider experiments
   H1\cite{Aid:1996bs} and 
   ZEUS\cite{Derrick:1995vq,Derrick:1996vw,Breitweg:1997ed,Breitweg:1999jy} 
   at HERA, where the $\gamma p$ CMS energy $W$ reaches values of up to 
   300~GeV.
   At large $W$ values, Regge phenomenology describes the diffractive 
   scattering as being mediated by the dominating exchange of the 
   Pomeron trajectory,  
   $\alpha_{\pom}(t) = \alpha_{\pom,0} + \alpha_{\pom}^{\prime}\cdot t$. The
   values $\alpha_{\pom,0} =1.08 - 1.10$
   and $\alpha_{\pom}^{\prime}=0.25$ GeV$^{-2}$ were determined from 
   global fits to hadron-hadron scattering 
   data\cite{Donnachie:1992ny,Cudell:1997ab,bib:dlslope}. 
   A first determination of the Pomeron trajectory with data from 
   exclusive \rhonull photoproduction was made by the ZEUS 
   collaboration\cite{Breitweg:1999jy}, using
   their own data at $W=94, 73$ and 71 GeV, as well as the H1 data at 
   $W=55$~GeV and the OMEGA\cite{Aston:1982hr}
   data at $W=8$ and 10~GeV. The result indicated a deviation of 
   $\alpha_{\pom}^{\prime}$ from the
   ``canonical'' value. 
   However, the inherent uncertainty in the combination of 
   cross sections from different experiments prohibited firm conclusions.
\par
   In the present analysis, data from a single experiment are
   for the first time
   used to extract the Pomeron trajectory for exclusive \rhonull
   photoproduction. For details of the analysis, see ref.11.

\section{Experimental procedure}

   The determination of $\alpha_{\pom}(t)$ from exclusive 
   \rhonull photoproduction in a single experiment
   requires large statistics over a wide range of $W$ and $t$, since the 
   $W$ dependence has to be determined in bins of $t$:
$
   \frac{\der \sigma\sub{\gamma p}\,(W)}{\der t}
   \propto
   \left ( \frac{W}{W\sub{0}} \right ) ^{4 (\alpha\sub\pom\,(t) - 1)}.
$
  \smallskip\par
   After the detector upgrade for the HERA-2 running period, H1 has taken the 
   new Fast Track Trigger (FTT)\cite{bib:ftt} into operation. 
   Combining charged track
   segments in the central drift chambers (CJC), the FTT 
   allows the triggering on final states with given total track charge and 
   multiplicity, and with a charged track $p_t$ threshold as low as 100 MeV. 
   In the first few
   months of stable operation of the FTT in summer 2005, 
   more than 271000 2-prong events with total charge 0 were collected in
   data corresponding to an integrated luminosity of 570 nb$^{-1}$. 
   More than 241000 of these events lie in the $\pi^+\pi^-$ 
   mass interval $0.6-1.1$ GeV.
   The background, mainly from production of
   other vector mesons ($\phi, \omega$ and $\rho^{\prime}$), is $< 2$\%.
   \par
   Since the triggering
   and event selection does not involve 
   the detection of the scattered electron, the modulus of the 
   squared momentum transfer at the electron vertex, 
   $Q^2$, is limited to $Q^2<4$~GeV$^2$  with $<Q^2>=0.01$~GeV$^2$. 
   The $W$ and $t$ ranges are $20<W<90$ GeV and $|t|<3$~GeV$^2$.   
   Since in the reaction   $e p \rightarrow e \rho Y$, where $Y$
   is either the proton or a proton dissociation system, the outgoing   
   proton system is not measured, 
   the variable $t$ is calculated from the total transverse 
   momentum $p_{t,\pi\pi}$ of
   the $\pi^+\pi^-$ system, $t = -p_{t,\pi\pi}^2$. Similarly,
   $W = \sqrt{2E_p(E_{\pi\pi} - p_{z,\pi\pi})}$,
   with $E_p$ being the nominal proton beam energy and $E_{\pi\pi}$ and 
   $p_{z,\pi\pi}$ the energy and 
   longitudinal momentum component of the $\pi\pi$ system.
\par
   Extensive Monte Carlo 
   studies were made,
   using the generator program DIFFVM\cite{List:1998jz}. 
   Signal events, i.e. events entering the cross section, are
   defined with the condition 
   $(M_Y^2 + Q^2)/(W^2+Q^2) < 0.01$, where $M_Y$ is the mass of the system $Y$.
   The MC events were reweighted in the variables $W$, $t$ and 
   $m_{\pi\pi}$, to obtain a good description of the experimental 
   distributions.
   \par
   The events are distributed in 
   twelve bins of $|t|$ and five to ten
   bins of $W$, altogether eighty bins, since the $W$ range varies with $|t|$. 
   In each $W,t$ bin, the
   average correction factor is $\sim 0.25$, 
   including acceptance, triggering and reconstruction efficiencies.
% \par
   The \rhonull 
   dipion mass spectrum was given by a Breit-Wigner function, skewed
   to the model of Ross-Stodolsky\cite{Ross:1965qa} \footnote{An analysis 
   using the S\"oding interference model\cite{Soding:1965nh} 
   gives completely equivalent results.}. 
   In each $W,t$ bin the number of \rhonull events was 
   obtained by first fitting to the data, 
   and then integrating the Breit-Wigner function
   from $m_{\pi\pi}$ threshold to 
   $m_{\rho^{\circ},0} + 5\Gamma_{\rho^{\circ},0}$, 
   where $m_{\rho^{\circ},0}$ and $\Gamma_{\rho^{\circ},0}$
   are the PDG\cite{Eidelman:2004wy} values. 
   The averages of the 80 fit values for
   $m_{\rhonull}$ and $\Gamma_{\rhonull}$ are very close to 
   the PDG values.
 \par
   The cross sections thus obtained
   include both the elastic and the proton 
   dissociative parts of the process.
   In order to separate these two components of the cross section, use is
   made of two subdetector systems in the ``forward'' direction of H1, 
   namely the 
   FTS (Forward Tagging System) and FMD (Forward Muon Detector). The
   dissociated proton system $Y$ produces hits in these detectors, 
   however, also elastic protons
   at larger values of $|t|$ may generate forward hits through secondaries. 
   The efficiencies of the elastic and dissociative 
   processes for forward hit generation were studied in detail with the 
   MC simulation, and the fractions
   of forward ``tagged'' (i.e. with hits in the forward subdetectors) 
   events in the MC simulation and in the data 
   were compared to
   obtain the number of elastic events in each $W,t$ bin. 
%
%   FIGURE 1
%
\smallskip
\par\noindent
\hspace*{-0.3cm}
\begin{minipage}[t]{5.8cm} 
\centerline{\epsfxsize=5.8cm\epsfbox{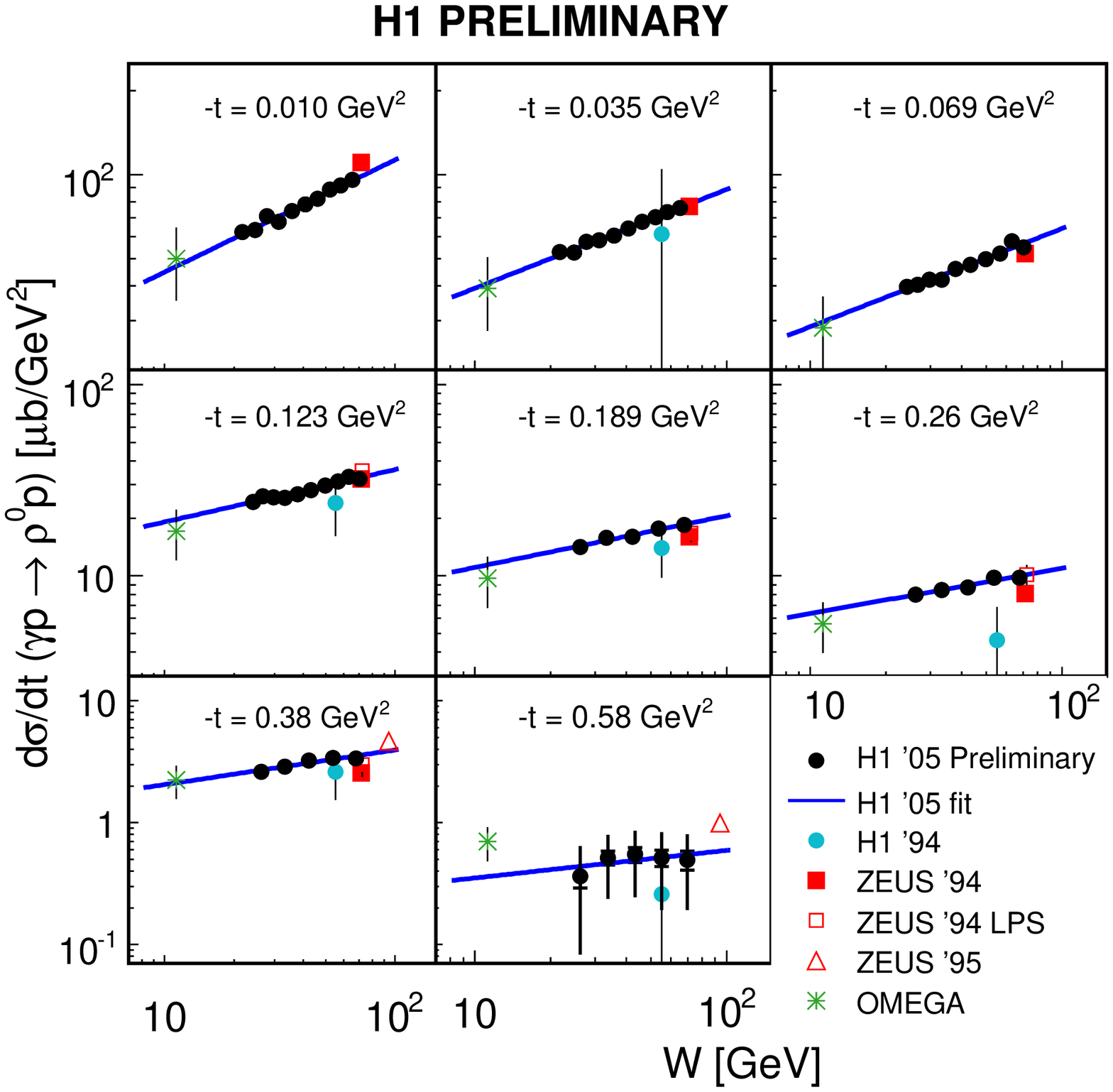}}
\end{minipage}   
\vspace*{-5.3cm}
\par\noindent
\hspace*{6cm}
\begin{minipage}[t]{5cm}
{\small Figure 1. \ \ 
$\gamma p$ cross sections for elastic  $\rho^{\circ}$ production 
as a function of $W$, in bins of $t$. The preliminary H1 data are compared to
previous results from the H1, ZEUS and OMEGA experiments. 
The curves show fits to the current H1 data alone, as discussed in the text. 
}
\end{minipage}
%\begin{figure}[ht]
%\centerline{\epsfxsize=5cm\epsfbox{H1prelim-06-011.fig6.eps}}   
%\caption{$\gamma p$ cross sections for elastic  $\rho^{\circ}$ production 
%as a function of $W$, in bins of $t$. The preliminary H1 data are compared to
%results from other experiments. The curves show fits to the current H1 data 
%alone, as discussed in the text. 
%    \label{figure1}}
%\end{figure}
\vspace*{1.3cm}
\section{Results}

  Figure 1 shows the resulting elastic \rhonull photoproduction cross section, 
  as a function of $W$ in 
   eight bins of $t$. 
In each $t$ bin the H1 data have been fitted to the function
  $\frac{\der \sigma\super{\gamma p}\,(W)}{\der t}
   = \frac{\der \sigma\super{\gamma p}\,(W_0)}{\der t}
     \left ( \frac {W}{W_0} \right ) ^{4 (\alpha - 1)}$, 
  with $W_0 = 37$ GeV. 
   Data and fits are compared to the previous measurements from
   ZEUS and H1, and to the lower $W$ data 
   from the OMEGA experiment. There is
   good agreement between the measurements, in particular the OMEGA data 
   agree well with the extrapolation of the fits to the H1 data.
\par
The fitted values of $\alpha$ are shown as a function of
  $t$ in Fig.2. The fit of the Pomeron trajectory 
   $\alpha_{\pom}(t) = \alpha_{\pom,0} + \alpha_{\pom}^{\prime}\cdot t$ 
   to the $\alpha$ values is also 
  shown, as is the global Pomeron trajectory of Donnachie and Landshoff. 
\par
The resulting preliminary values of the fit parameters are  
$\alpha_{\pom,0}=\OurAlphaZero$
and $\alpha_{\pom}^{\prime}=\OurAlphaPrime$.
While the intercept agrees well with the expected value range,
  the slope $\alpha_{\pom}^{\prime}$ differs significantly from the
  canonical value 0.25 GeV$^{-2}$. This indicates 
  that a model, in which the exchange of 
  one global, linear Pomeron trajectory explains
  all diffractive scattering at high energies, is too simple.
%
%   FIGURE 2
%
\par\noindent
\hspace*{-0.3cm}
\begin{minipage}[t]{5.7cm}
\centerline{\epsfxsize=5cm\epsfbox{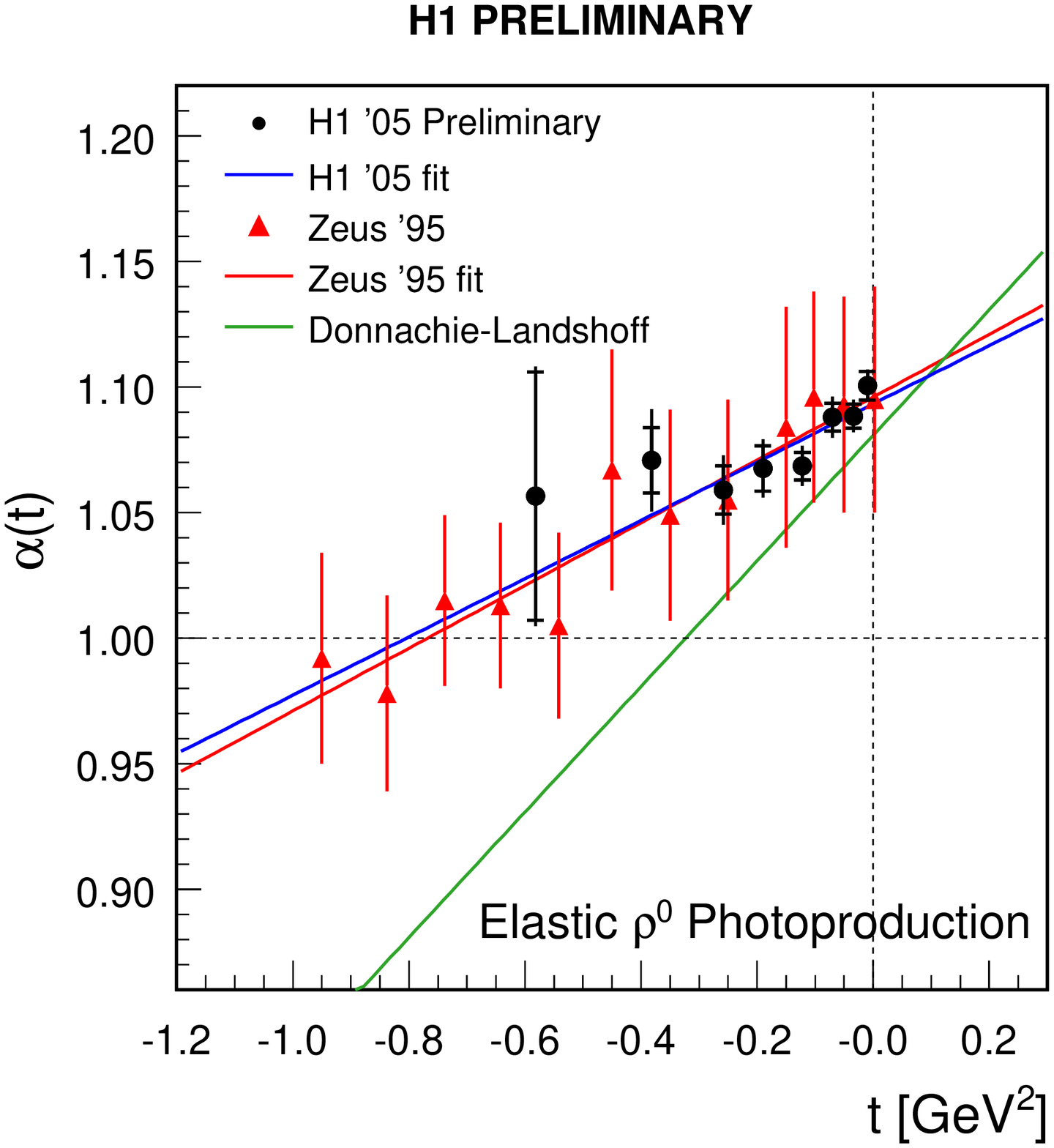}}
\end{minipage}
\vspace*{-5cm}
\par\noindent
\hspace*{5.7cm}
\begin{minipage}[t]{5.5cm}   
{\small Figure 2. \ \ 
The measured values of the parameter $\alpha$ for elastic 
\rhonull photoproduction, together with a fit of the 
Pomeron trajectory of the form
$\alpha\sub{\pom}\,(t) = \alpha\sub{\pom,0} + \alpha'\sub{\pom}\cdot t$.
The previous result, obtained using data from the ZEUS, 
H1 and OMEGA experiments, is also shown, as is the parametrisation from 
Donnachie and Landshoff.
}
\end{minipage} 
%\begin{figure}[ht]
%\centerline{\epsfxsize=5cm\epsfbox{H1prelim-06-011.fig10.eps}}   
%\caption{The measured values of the parameter $\alpha$ for elastic 
%\rhonull photoproduction, together with a fit to the 
%Pomeron trajectory of the form
%$\alpha\sub{\pom}\,(t) = \alpha\sub{\pom,0} + \alpha'\sub{\pom}\cdot t$.
%The previous result, obtained using data from the ZEUS, 
%H1 and OMEGA experiments, is also shown, as is the parametrisation from 
%Donnachie and Landshoff.    \label{figure2}}.
%\end{figure}
\vspace*{0.9cm}
\par
The present result 
  agrees well with a previous result which was obtained using ZEUS, H1
  and OMEGA data at $W=94,73,71,55,10$ and 8~GeV. 
  This previous result is also shown in Fig.2.

\end{document}